\documentclass[journal=jpclcd,manuscript=letter]{achemso}

\usepackage{amsmath}
\usepackage{tabularx}
\usepackage{multirow}
\usepackage{subfigure}
\usepackage{ulem}
\usepackage{caption}

\usepackage{xcolor}

\definecolor{ykred}{rgb}{1.0, 0.0, 0.0}

\definecolor{sbblue}{rgb}{0.0, 0.0, 1.0}

\definecolor{jxblue}{rgb}{0.8, 0.5, 0.8}

\definecolor{rzblue}{rgb}{1.0, 0.0, 1.0}

\definecolor{modred}{rgb}{1, 0.5, 0}

\author{Sampreeti Bhattacharya}
\affiliation{Department of Chemistry, University of North Carolina at Chapel Hill, Chapel Hill, NC, 27514, USA}

\author{Jianhang Xu}
\affiliation{Department of Chemistry, University of North Carolina at Chapel Hill, Chapel Hill, NC, 27514, USA}
\author{Ruiyi Zhou}
\affiliation{Department of Chemistry, University of North Carolina at Chapel Hill, Chapel Hill, NC, 27514, USA}

\author{Yosuke Kanai}
\affiliation{Department of Chemistry, University of North Carolina at Chapel Hill, Chapel Hill, NC, 27514, USA}
\altaffiliation{Department of Physics and Astronomy, University of North Carolina at Chapel Hill, Chapel Hill, NC, 27514, USA}
 \email{Corresponding author: ykanai@live.unc.edu}

\title{Proton Quantum Effects on Electronic Excitation in Hydrogen-bonded Organic Solid: A First-Principles Green's Function Theory Study} 

\begin{document}
\newcommand{\kpe}{\mathbf{k}\!\cdot\!\mathbf{p}\,}
\newcommand{\Kpe}{\mathbf{K}\!\cdot\!\mathbf{p}\,}
\newcommand{\bfr}{ {\bf r}} 
\newcommand{\bfrp}{ {\bf r'}} 
\newcommand{\tp}{ {t^\prime}} 
\newcommand{\bfrpp}{ {\bf r^{\prime\prime}}} 
\newcommand{\bfR}{ {\bf R}} 
\newcommand{\bfRp}{ {\bf R'}} 
\newcommand{\bfG}{ {\bf G}} 
\newcommand{\bfGp}{ {\bf G'}} 
\newcommand{\bfzero}{ {\bf 0}} 
\newcommand{\bfq}{ {\bf q}} 
\newcommand{\bfp}{ {\bf p}} 
\newcommand{\bfk}{ {\bf k}} 
\newcommand{\dv}{ {\Delta \hat{v}}} 
\newcommand{\sigmap}{\sigma^\prime} 
\newcommand{\omegap}{\omega^\prime} 
\newcommand{\omegapp}{\omega^{\prime\prime}} 
\newcommand{\bracketm}[1]{\ensuremath{\langle #1   \rangle}}
\newcommand{\bracketw}[2]{\ensuremath{\langle #1 | #2  \rangle}}
\newcommand{\bracket}[3]{\ensuremath{\langle #1 | #2 | #3 \rangle}}
\newcommand{\ket}[1]{\ensuremath{| #1 \rangle}}
\newcommand{\GnWn}{\ensuremath{G_\text{0}W_\text{0}}\,}
\newcommand{\tn}[1]{\textnormal{#1}}
\newcommand{\f}[1]{\footnotemark[#1]}
\newcommand{\mc}[2]{\multicolumn{1}{#1}{#2}}
\newcommand{\mcs}[3]{\multicolumn{#1}{#2}{#3}}
\newcommand{\mcc}[1]{\multicolumn{1}{c}{#1}}
\newcommand{\refcite}[1]{Ref.~\cite{#1}} 
\newcommand{\refsec}[1]{Sec.~\ref{#1}} 
\newcommand\opd{d}
\newcommand\im{i}
\def\bra#1{\mathinner{\langle{#1}|}}
\def\ket#1{\mathinner{|{#1}\rangle}}
\newcommand{\braket}[2]{\langle #1|#2\rangle}
\def\Bra#1{\left<#1\right|}
\def\Ket#1{\left|#1\right>}
\def\onlinecite{\cite[left=,right=]}

\newcommand{\fatr}{\mathbf{r}}

\newcommand{\vect}[1]{\boldsymbol{#1}}
\begin{tocentry}
\centering
\includegraphics[height=5cm,width=5cm]{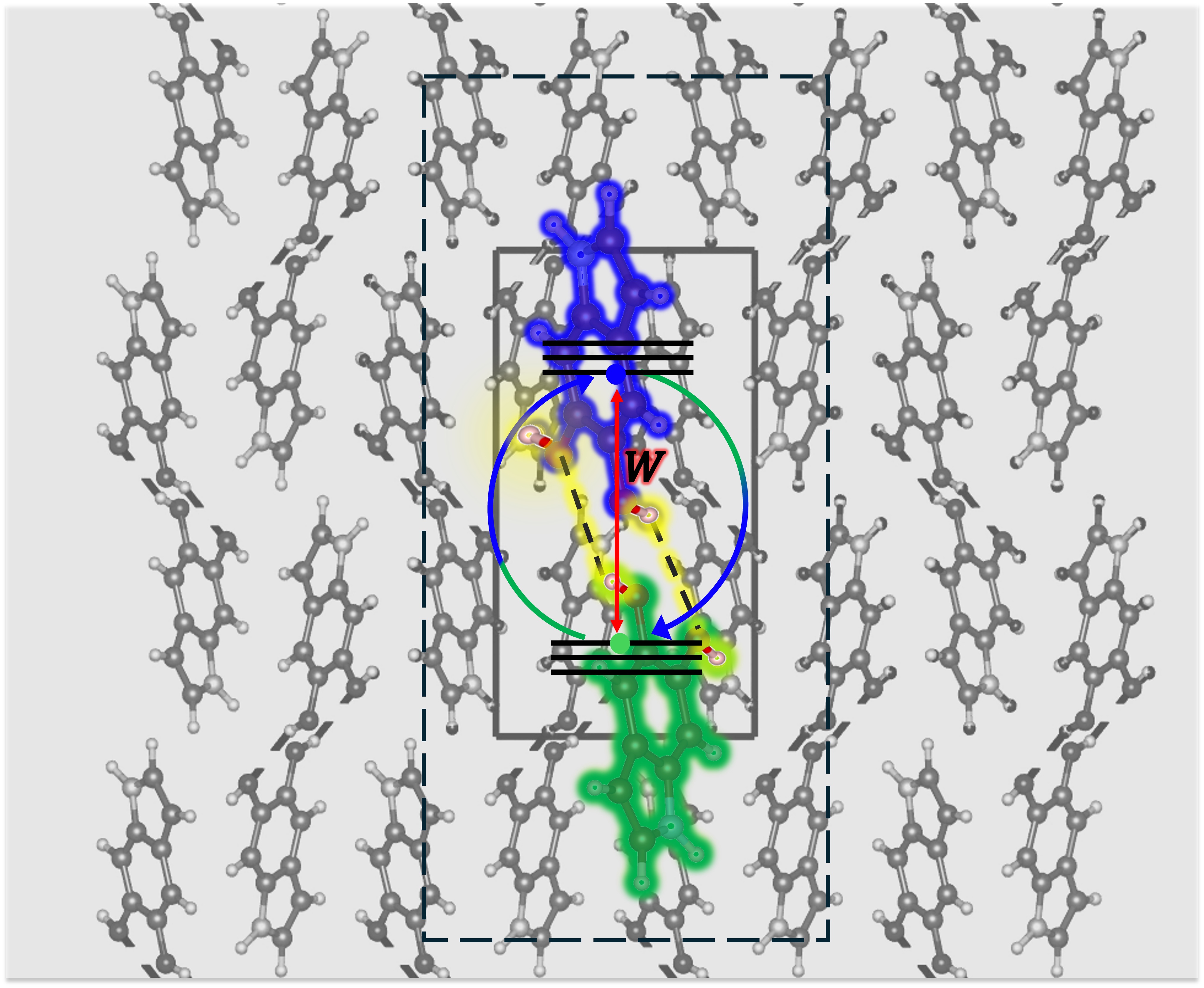}
\end{tocentry}
\begin{abstract}
 Nuclear quantum effects of protons on electronic excitations in hydrogen-bonded organic materials remains underexplored.
In theoretical studies, 
modeling excitons in these extended systems is particularly difficult because they tend to have a large exciton binding energy and sometimes exhibit charge transfer character.
We demonstrate how first-principles Green's function theory combined with the nuclear-electronic orbital method enables us to examine the nature of
excitons in a prototypical organic solid of eumelanin, for which the extensive hydrogen bonds have been proposed to facilitate the formation of delocalized excitons.
We investigate how the quantization of protons impacts electronic excitations.
We discuss the extent to which the resulting proton quantum effects can be described as being derived from structure and how they induce molecular-level anisotropy for the excitons in the organic solid. 
\end{abstract}

\noindent
Recent studies continue to highlight important roles of nuclear quantum effects (NQEs) in various chemical processes from proton-coupled electron transfer, proton dissociation, to thermodynamic and optical properties of water
\cite{wang2014quantum,clark2019opposing,tang2021nuclear,eltareb2021nuclear,li2022static,berrens2024nuclear,cui2025theoretical}. 
Quantum mechanical description of light nuclei like protons is increasingly recognized as being essential, especially in condensed matter systems with extensive hydrogen bond networks \cite{berrens2024nuclear,cui2025theoretical,tang2025optical,rashmi2025resolving}. 
This is particularly relevant because hydrogen bonds also represent a key interatomic interaction  in organic materials and in other soft matter systems, determining the macroscopic properties of supramolecular functional materials and metal-organic frameworks \cite{hanikel2021evolution}. 
For instance, in liquid crystal polymer systems, hydrogen bonds enable stimuli responsive behavior, self healing, and recyclability \cite{lugger2021hydrogen}. 
In theoretical works, first-principles molecular dynamics (FPMD) simulations, with path-integral (PI) methods \cite{feynman1965path,feynman1965quantum,feynman1979path,topaler1996path,herrero2014path,ananth2022path,poltavsky2016modeling}, 
have shown that the quantum-mechanical treatment of protons is essential for accurately describing the dynamics and structure of liquid water\cite{paesani2007quantum,paesani2010nuclear}, molecule-solvent interactions  \cite{zuehlsdorff2018unraveling} etc. 
Interestingly, the NQEs are often discussed in terms of having competing effects of weakening and strengthening hydrogen bonds in water \cite{paesani2007quantum,paesani2010nuclear,ceriotti2016nuclear,marsalek2017quantum,yao2020temperature,wang2016simulating,yao2021nuclear}. 
This NQE induced modulation is not unique to water;  and has also  been observed across a wide variety of hydrogen-bonded systems\cite{cha1989hydrogen,li2011quantum,sauceda2021dynamical}.
However, while the effect of nuclear quantization on structural and dynamical properties is well documented for both  water (see e.g.\cite{morrone2008nuclear,fritsch2014nuclear,ceriotti2016nuclear,markland2018nuclear,yao2021nuclear}) and organic solids \cite{cazorla2017simulation,markland2018nuclear}, their impact on electronic excitation remains largely unexplored. 
Among the limited studies that exist, \citeauthor{tang2021nuclear} have theoretically investigated the role of NQEs in the context of near-edge X-ray absorption fine structure spectra of water\cite{tang2021nuclear}.


In most electronic structure calculations,
atomic nuclei, including protons, are modeled as classical point charges by employing the Born-Oppenheimer (BO) approximation.  
In recent years, the nuclear-electronic orbital (NEO) method has emerged
as a particularly convenient approach for taking into account nuclear quantum effects within electronic structure theory\cite{webb2002multiconfigurational,iordanov2003vibrational,pak2004electron,swalina2006explicit,hammes2021nuclear,xu2022nuclear}.
The NEO method enables us to ``quantize" selected atomic nuclei on an equal footing with electrons, making the approach particularly convenient for studying complex condensed-phase matter. 
This approach has been successfully used, for example, for implementing multicomponent density functional theory (DFT) \cite{capitani1982non,kreibich2001multicomponent} based on the Kohn-Sham formalism such that more than one type of particles can be quantized (e.g., electrons and protons), and the correlation functional allows us transcend the BO approximation  \cite{chakraborty2008development,udagawa2014electron,yang2017development}.

\begin{figure}[htbp]
    \captionsetup{justification= centerlast}
    \centering
       \subfigure[]{\label{fig:geom_herr}\includegraphics[width=75mm,height=60mm]{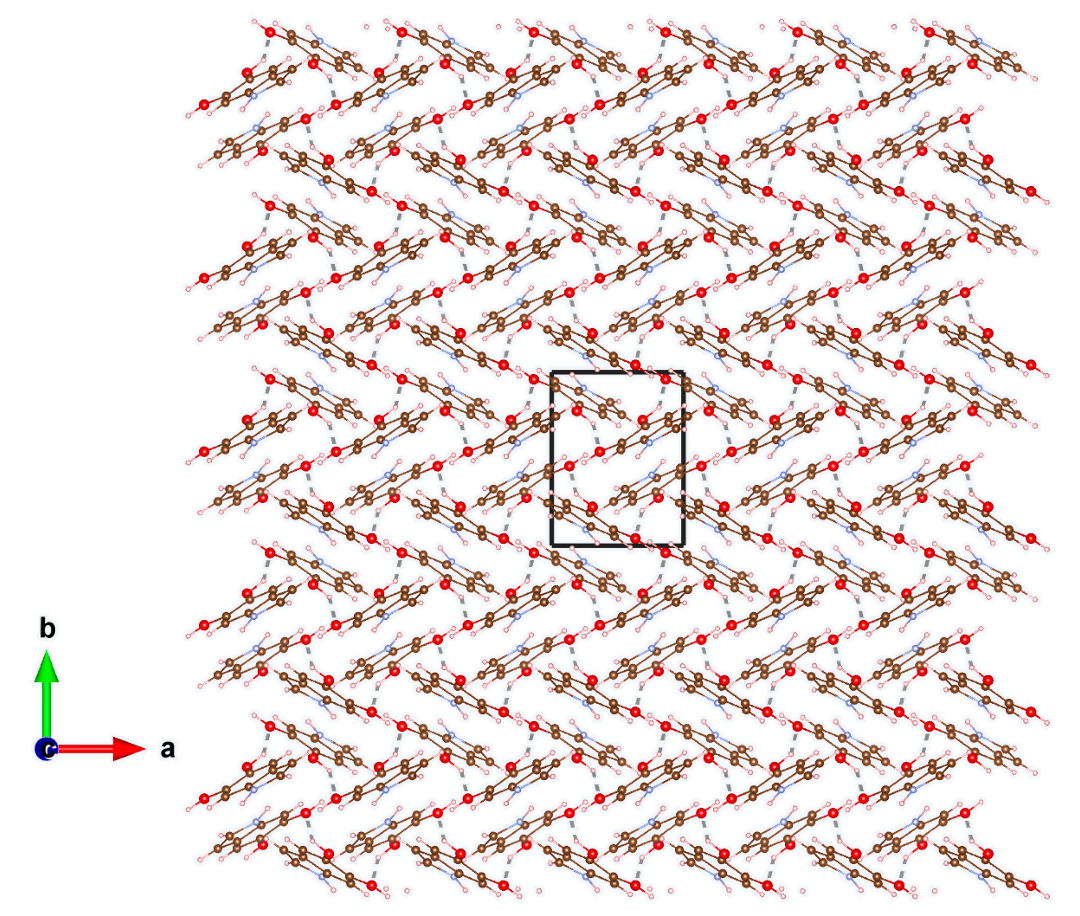}}
  \hspace{2mm}
\subfigure[]{\label{fig:geom_1}\includegraphics[width=82mm,height=63mm]{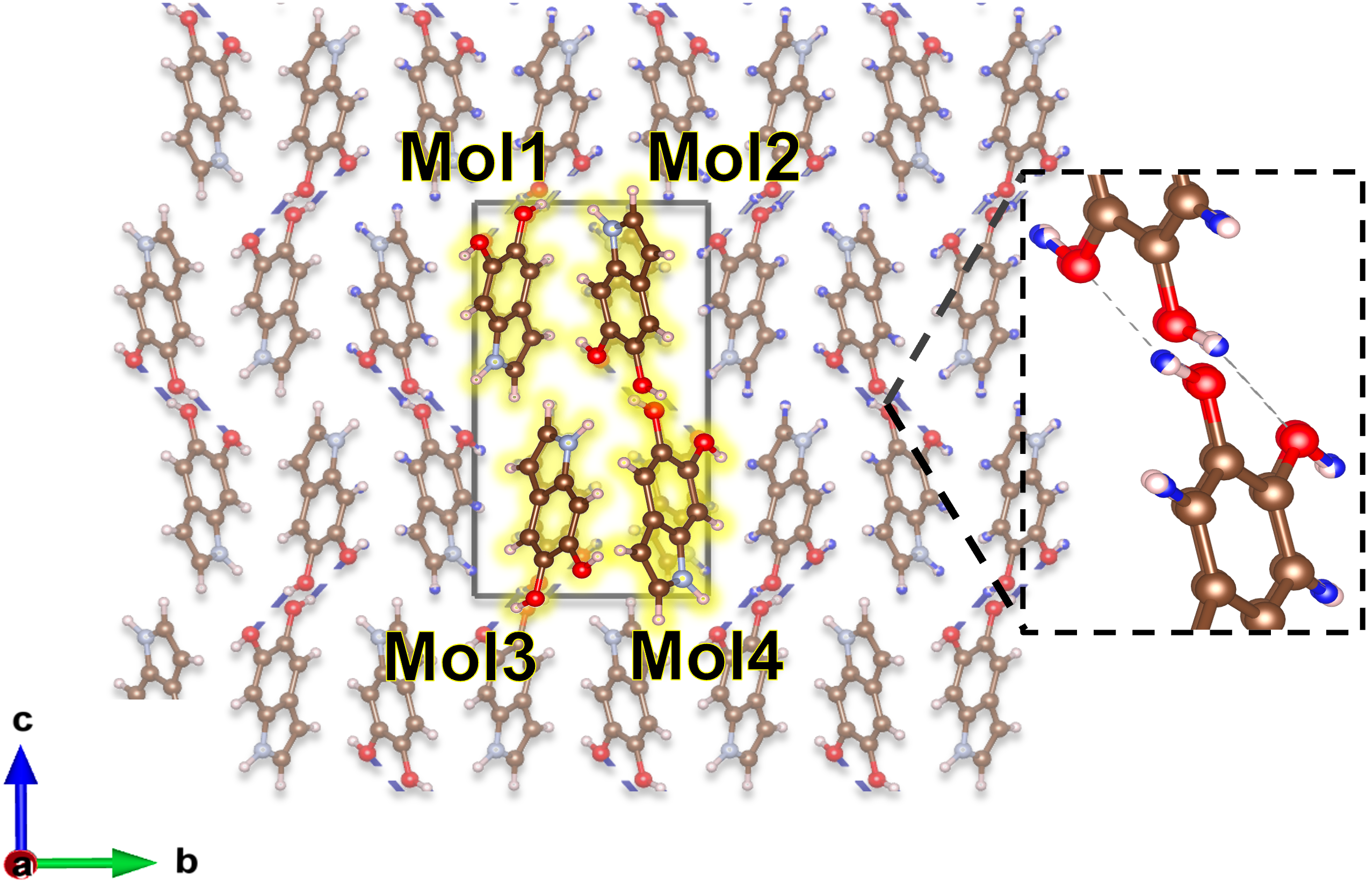}}
\caption{(a) DHI crystal structure with the unit cell shown as a black box. The organic solid has the helical packing with the herringbone pattern in the crystal. 
(b) Monomer units in the unit cell are highlighted in yellow. Inset shows the position expectation values of quantum protons are highlighted in blue, overlaid on classical proton positions shown in pink.  
Mol1-Mol3 and Mol2-Mol4 monomer pairs are hydrogen-bonded. }
\label{fig:hbonds_figure}
\end{figure}

The Bethe-Salpeter Equation (BSE), based on the $GW$  approximation ($G$ stands for Green’s function and $W$ for the screened Coulomb interaction)  within the many-body Green’s function perturbation theory, 
is widely considered as one of the most successful first-principles electronic structure methods for modeling excited states of extended systems, overcoming key shortcomings of the linear-response time-dependent DFT method with standard exchange-correlation (XC) approximations\cite{hankeManyParticleEffectsOptical1979, salpeterRelativisticEquationBoundState1951,strinati1982dynamical,hybertsen1986electron,rohlfing2000electron,golze2019gw,bhattacharya2024bse,abbott2025roadmap}.
The BSE method effectively solves for the particle-hole wavefunction associated with individual excitons, building on the quasiparticle (QP) description of the charged excited states, which is obtained from the $GW$ method\cite{hedinNewMethodCalculating1965,hybertsen1985first,onida2002electronic}.  
The success of the BSE@$GW$ approach can be partially attributed to the screened Coulomb interaction kernel for describing the particle-hole interaction while the $GW$ method already takes into account the quasi-particle energies for the charged excited states. 
On the other hand, the XC kernel in linear-response TDDFT would need to describe both of these effects simultaneously as it lacks the single-particle states renormalized by the self-energy\cite{ullrich2012time}.
Investigations into the nuclear quantum effects (NQEs) on electronic excitation are much more limited in the literature compared to their ground-state counterparts. 
In order to take into account the NQEs, the path integral (PI) formalism is widely used, such that the quantum delocalization of atomic nuclei is modeled by discrete beads in imaginary time\cite{feynman1979path,makri2015quantum,althorpe2021path,ananth2022path,marx1996ab}. 
Individual PI beads have their own electronic structure that is obtained for the given coordinates of classical atomic nuclei. 
Such path-integral simulations have been used together with BSE and $GW$ calculations to study the NQEs on electronic excitation of liquid water by Wu and co-workers \cite{tang2025optical}.
Instead, we here investigate the NQEs on the electronic excitation of a hydrogen-bonded organic crystalline solid by using the NEO method in the BSE@$GW$ calculation.
Delocalization of excitons (particle-hole pairs) in bifurcated hydrogen-bonded dimers was proposed 
to explain the photo-function of eumelanin composed of 
5,6 dihydroxyindole (DHI)
\cite{sasikumar2022exciton}, and we study the proton quantum effects on excitons in this prototypical organic solid with extensive hydrogen bonds.
Crystalline DHI, as shown in Figure \ref{fig:geom_herr}, exhibits an extensive helical hydrogen-bonded packing along with $\pi$-$\pi$ stacking, and these structural features are believed to be important in determining its excited state properties \cite{chen2017polydopamine,ghosh2021computational,sasikumar2022exciton}.
Motivated by these works, we examine how NQEs, proton quantum effects in particular, impact the electronic excitation of this crystalline organic solid \cite{eric2023manifestation,fresch2023role,micillo2017eumelanin}, using first‑principles many‑body Green’s function calculations.



We use the geometry as obtained from X‑ray diffraction (XRD) measurements \cite{sasikumar2022exciton}, and 
the unit cell contains four monomer units that are extensively hydrogen-bonded.
For concise discussion, we adopt the following nomenclature to refer to three different types of calculations. 
Standard DFT calculations are referred to as \textit{Std}, and the calculations with quantized protons using NEO method are termed as \textit{NEO}. 
In order to quantify the extent to which the NQEs on electronic excitations are due to the geometry-induced effects (as opposed to its direct role in modifying the electronic Hamiltonian), we further perform additional standard DFT calculations with protons modeled as classical point charges \textit{but} with their locations given by the position expectation value of quantum protons from the \textit{NEO} calculation.
These calculations are referred to as \textit{Std:QGeom}.  For each calculation, only the hydrogen positions are relaxed while other types of atoms are in the positions as obtained using XRD \cite{sasikumar2022exciton}.
Further computational details are discussed in the \textit{Theoretical/Computational Details} section and in Supporting Information. 
In the equilibrium (0K) geometry for NEO-DFT (\textit{NEO}) calculations, the expectation values of the position operator for quantum protons are shifted by as much as $\sim$0.015 {\AA} from the hydrogen atom positions in the standard DFT (\textit{Std}) calculation. 
Table S1 of the Supporting Information summarizes the C-H, O-H, and hydrogen-bond distances. 
The differences between the \textit{NEO/Std:Qgeom} and \textit{Std} calculations are within the range of PIMD and DFT values  previously reported by Miranda et al. for both O-H and hydrogen-bond distances\cite{10.1063/5.0279956}. 
The average C-H bond length increases by 0.03 {\AA} in the \textit{NEO} calculation, compared to the \textit{Std} calculation. 
Unlike C-H bond distances, the average O-H bond length decreases by 0.02 {\AA} when protons are quantized, while the average hydrogen-bond distance shortens to 2.04 {\AA} from 2.07 {\AA}.

\begin{figure}[htbp]
 \captionsetup{justification= centerlast}
\centering
\subfigure[]
{\label{GW_dos}\includegraphics[width=53mm,height=42mm]{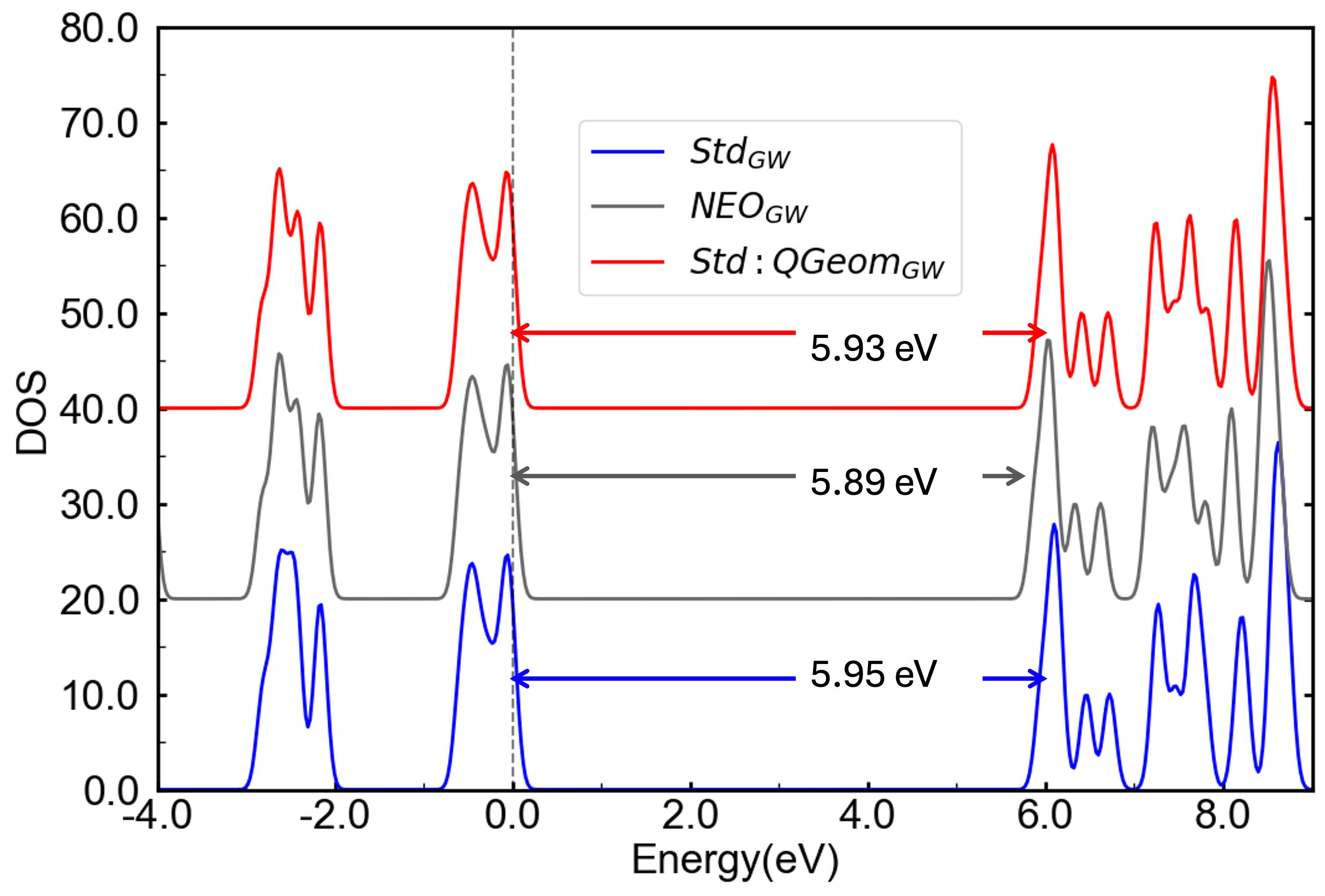}}
\subfigure[]{\label{fig:std_neoall}\includegraphics[width=53mm,height=42mm]{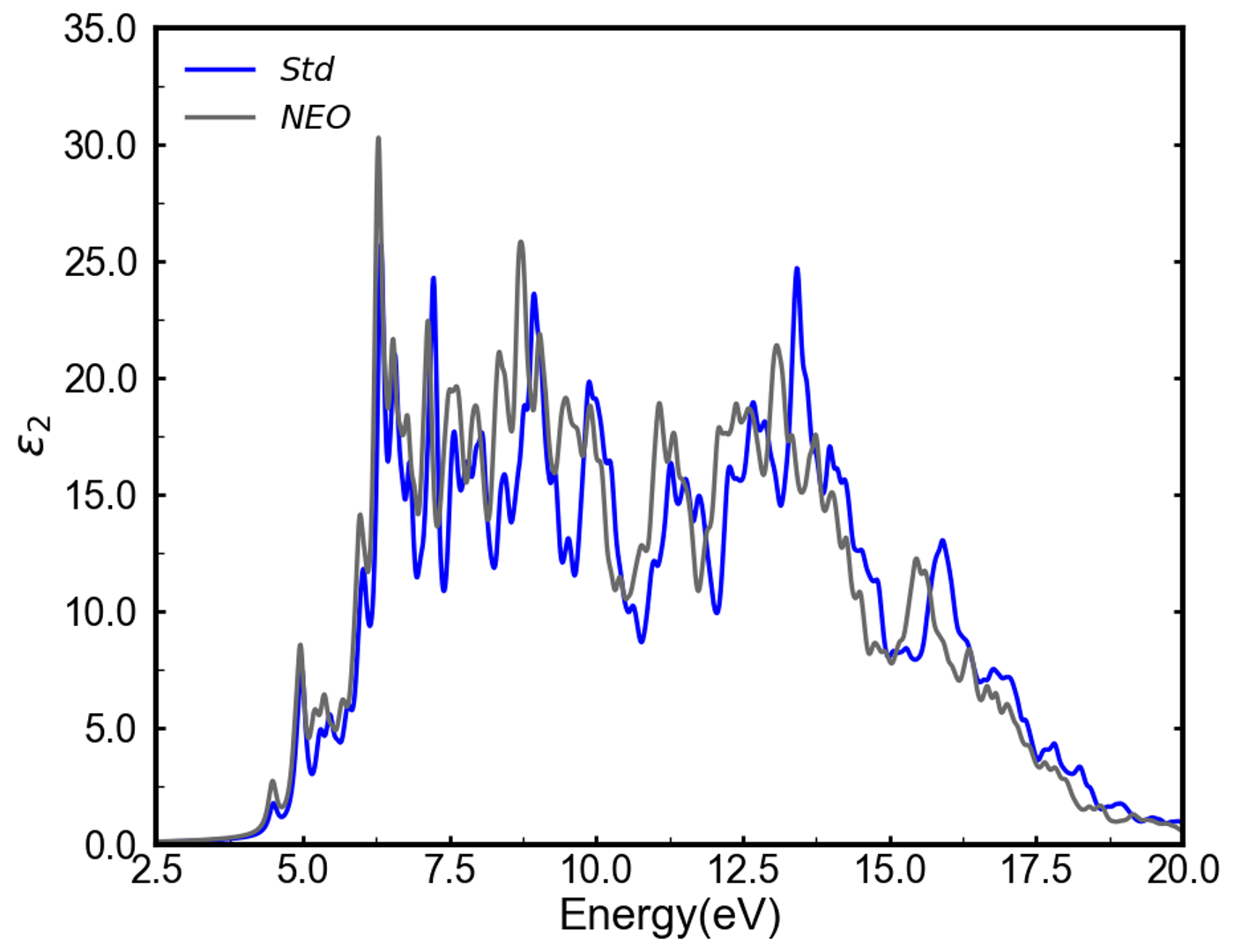}}
 \subfigure[]{\label{fig:qgeom_neoall}\includegraphics[width=53mm,height=42mm]{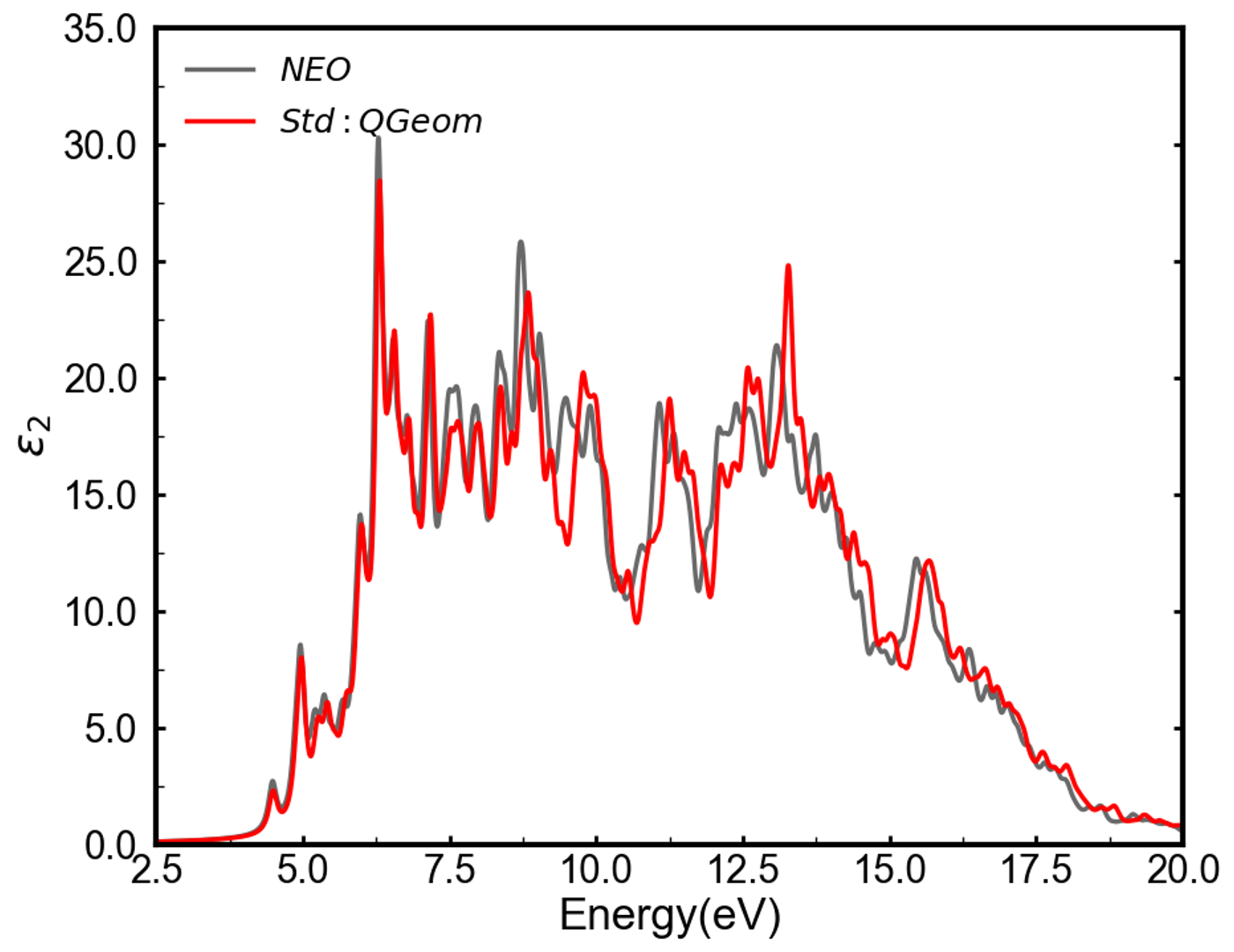}}
\caption{(a) Quasi-particle (QP) density of states (DOS) obtained from $GW$ calculations. 
The x-axis shows QP energy in eV, and the DOS is aligned with respect to the HOMO as a reference (0.0 eV). Gaussian broadening with a FWHM value of 0.025 eV was used. 
(b) Comparison of the optical absorption spectra for \textit{Std} (in blue) and \textit{NEO} (in grey) calculations obtained using BSE@$GW$ calculations.
(c) Comparison of the optical absorption spectra for \textit{NEO} (in grey) and \textit{Std:QGeom} (in red) calculations obtained using BSE@$GW$ calculations.
The x-axis shows the excitation energy in eV, and the optical absorption spectrum for the solids is given by the imaginary part of dielectric function in the y-axis.}
\label{fig:opt}
\end{figure}

Quasi-particle (QP) (i.e. charged excitation) energies and the corresponding density of states (DOS) are calculated using the
$GW$ method and shown in Figure \ref{GW_dos} (Corresponding DOS from DFT is shown in Figure S3 in the Supporting Information). 
By incorporating the proton quantum effects, the QP energy gap noticeably decreases by more than 0.05 eV from 5.95 eV (\textit{Std}) to 5.89 eV (\textit{NEO}).
For the energy range below -2.0 eV and above 7.0 eV, one observes noticeable differences between the \textit{Std} and the \textit{NEO} calculations. 
Interestingly, the \textit{Std:QGeom} calculation shows many of the same DOS features as the NEO calculation, indicating that the proton quantum effects here are mostly geometry-derived.
We now examine the effects of proton quantization on excited-state properties using the BSE@$GW$ method in the NEO framework. We compute electronically excited states within the spin singlet manifold, and we assume that the Franck-Condon principle is also valid for quantum protons in the sense that the electronic transitions are not coupled with the nuclear transitions because of the large difference in the relevant frequency range. 
In terms of the Green's function, this amounts to neglecting the phonon contribution to the quantum proton component in the self-energy. 
Without the proton quantum effects, the first/lowest excitation energy is 4.49 eV (\textit{Std}) while quantizing protons changes the excitation energy to 4.48 eV (\textit{NEO}). Essentially, no changes are observed for the optical energy gap. 
If the quantum proton position expectation values are used for classical proton positions, the calculation yields the excitation energy of 4.48 eV (\textit{Std:QGeom}). 
Another particularly important excited-state property for organic solids is the exciton binding energy, which quantifies how strongly the excited electron and hole are bound as a particle-hole pair and represents the energetic cost for charge separation\cite{kanai2010theory}.
The exciton binding energy, $E_b$, is formally defined as the difference between the neutral excitation energy (i.e., optical energy gap) and the quasi-particle energy gap (i.e., $E_b=E_{gap}^{QP}-E_{gap}^{Opt.}$). 
As summarized in Table \ref{Tab:exciton}, quantizing protons decreases the exciton binding energy from 1.46 eV (\textit{Std}) to 1.41 eV (\textit{NEO}).  
Accounting only for the geometry-derived effect of quantum protons as in the \textit{Std:QGeom} calculation, the exciton binding energy is 1.44 eV. 
Comparing the \textit{Std} and \textit{NEO} calculations in Table \ref{Tab:exciton}, 
one notes that the proton quantum effects on the exciton binding energy largely stems from the quasi-particle energy gap, $E_{gap}^{QP}$ and not from the optical gap. 
\begin{table}[]
\centering
\begin{tabular}{c|c|c|c}
\hline \hline
 & QP Gap (eV) & Optical Gap (eV) & Exciton Binding Energy (eV) \\ \hline
\textit{Std }      & 5.95   & 4.49             & 1.46            \\ \hline
\textit{NEO  }     & 5.89   & 4.48             & 1.41            \\ \hline
\textit{Std:QGeom} & 5.93   & 4.49             & 1.44            \\ \hline
\end{tabular}
\caption{Quasi-particle (QP) gap, optical gap, and exciton binding energy ($E_b=E_{gap}^{QP}-E^{Opt}_{gap}$) for  $Std$, $NEO$, and $Std:QGeom$ calculations.  }
\label{Tab:exciton}
\end{table}

Optical absorption spectra calculated using the BSE@$GW$ are shown in Figure \ref{fig:opt}. 
For extended systems like organic crystals, the optical absorption spectrum corresponds to the imaginary part of the dielectric function, $\varepsilon_2(\omega)$.
Qualitatively, quantizing protons does not significantly change the overall shape of the optical absorption spectrum as can be seen by comparing the \textit{NEO} and \textit{Std} calculations in Figure \ref{fig:std_neoall}. 
The most notable change is that the proton quantum effects red-shift the spectrum above 10 eV.
The peak heights of some prominent excitations at around 4.5, 5.0, and 6.2 eV also become noticeably more pronounced when protons are quantized.
The question, again, is to what extent the observed changes are due to the geometric effect from having quantum protons \cite{tang2021nuclear}.
As can be seen in Figure \ref{fig:qgeom_neoall}, 
both the \textit{NEO} and \textit{Std:QGeom} calculations show nearly identical peak positions with only minor differences in $\varepsilon_2$ intensities, 
indicating that the proton quantum effects here are mostly geometric effects.
It is more of an indirect quantum effect that is due to the  shifting of the centers of quantum proton densities away from classical proton positions.

\begin{figure}[htbp]
    \captionsetup{justification= centerlast}
    \centering
\includegraphics[width=155mm,height=90mm]{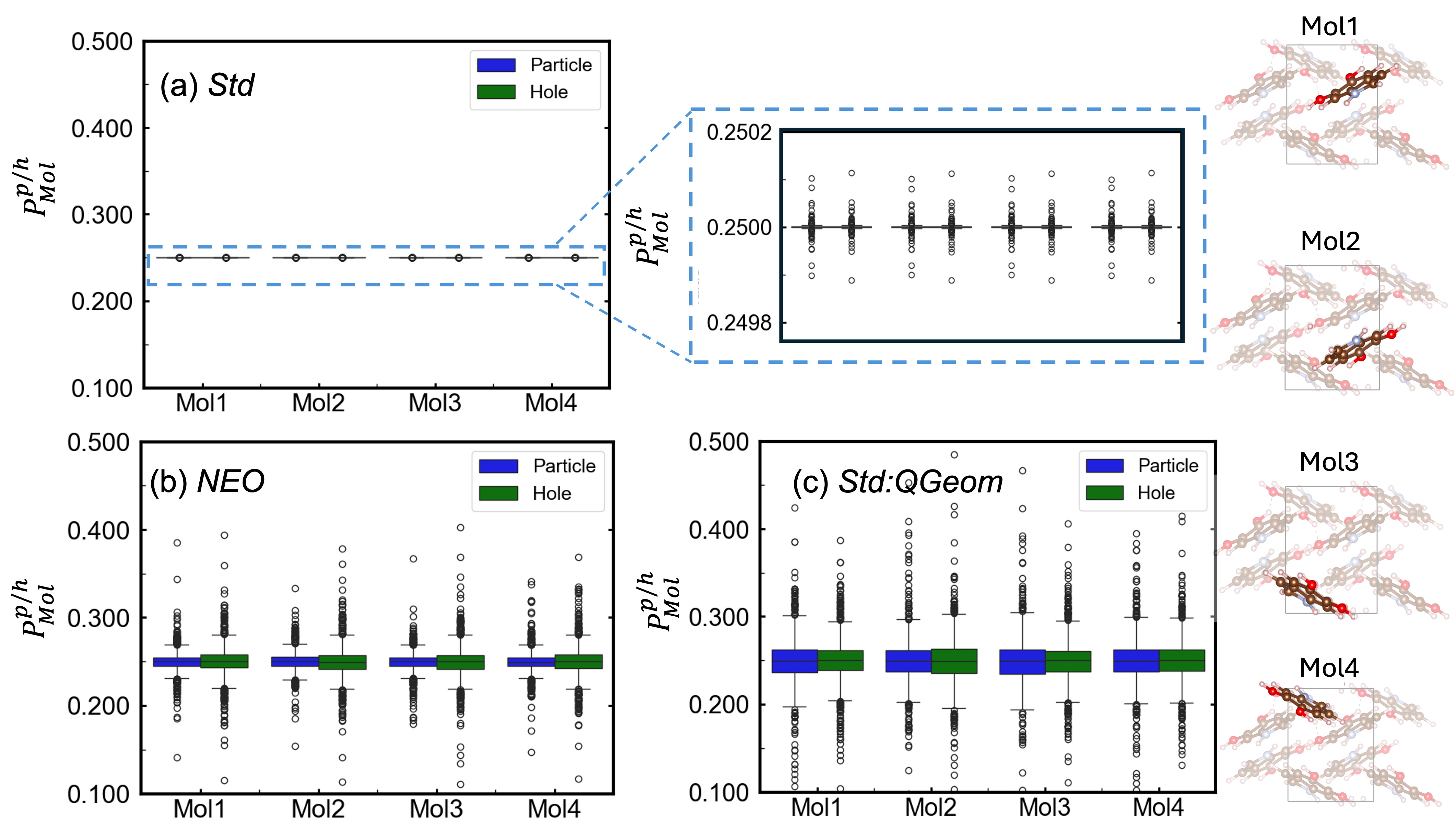}
 \caption{ 
 Distribution of Mulliken exciton populations for individual monomers, $P^{p/h}_{Mol}$, for (a) \textit{Std}, (b) \textit{NEO}, and (c) \textit{Std:QGeom} calculations. 
 The Mulliken exciton populations 
 are shown for the 
 particle and hole of the excitons in blue and green, respectively.
 Each circle indicates  $P^{p/h}_{Mol}$ for individual excited state $n$.
In (a)-(c), the box heights correspond to the standard deviation of the Mulliken population distributions (see Eq. \ref{eq:spread}). 
In (a), the inset shows a magnified y-axis to highlight the standard deviation for \textit{Std} calculation. 
The horizontal lines passing through the boxes shows the average value given by $\mu^{p/h}_{P_{Mol}}$. 
Deviations of the values from 0.25 indicate that the excitons are spatially more heterogeneously distributed among the four monomers in the unit cell. }
\label{fig:sd_delocalization}
 \end{figure}

\par 
Although the optical absorption spectra remain qualitatively similar even when proton quantum effects are taken into account (Figure \ref{fig:opt}),
the oscillator strengths for individual excitations show some differences (see Figure~S4 in Supporting Information), prompting us to examine how the proton quantization impacts the nature of individual electronic excitations.
Hydrogen bonds have been proposed to mediate exciton delocalization in the DHI crystal in the literature\cite{sasikumar2022exciton}, and proton quantum effects are likely important in hydrogen bonds.
Additionally, the theoretical work by Kundu and Galli has shown that the zero-point vibrational motion leads to dynamic Jahn-Teller-like lifting of the electronic degeneracy \cite{kundu2024quantum}, and such NQEs-induced symmetry breaking might also be present for excitons in these extended systems with extensive hydrogen bonds. 

Exciton probability density in individual excited states can be obtained in the BSE@$GW$ method (see Eq. \ref{eq:prob_density} 
in \textit{Theoretical/Computational Details} section). 
Within the two-particle Green's function description, individual excited states are represented by a single exciton quasi-particle or a particle-hole pair. 
The exciton probability density describes the conditional probability between the particle (i.e. excited electron) and the hole, and it is quite nontrivial for interpretation.
We can instead introduce the mean probability density for the particle by integrating out the hole degree of freedom for each excited state (Eq. \ref{eq:e_density}), and vice-versa for the hole (Eq. \ref{eq:h_density}). 
For molecular crystals with distinct monomer units like DHI, it is instructive to quantify how the exciton probability density is distributed among the monomers in individual excited states.
To quantify such anisotropic features,  Mulliken exciton populations for the particle and hole are calculated for individual monomers as discussed in \textit{Theoretical/Computational Details} section (see Eqs. \ref{eq:density_matrix_particle_mo}, \ref{eq:density_matrix_hole_mo}, and \ref{eq:Mol_decom}).
Figure \ref{fig:sd_delocalization} shows the distributions of the Mulliken exciton populations on individual monomer units 
for the lowest 200 excited states, covering 4.5-8.2 eV in the excitation energy.
The Mulliken exciton populations, $P^{p/h}_{Mol, n}$ ($Mol=1\sim4$ and $n=1\sim200$), are shown for each monomer using box-plots; the height of the boxes represents the standard deviation. 
When protons are modeled as classical point charges, the Mulliken exciton populations are essentially the same on all monomers for the particle and hole in all the excited states (i.e. $P^{p/h}_{Mol, 1}\approx0.25$). 
This is perhaps not surprising given the highly symmetric nature of the DHI crystal in terms of atom positions. 
\begin{figure}[htbp]
\captionsetup{justification= centerlast}
\centering
 \subfigure[]{\label{fig:P_mol_diff_el}\includegraphics[width=65mm,height=80mm]{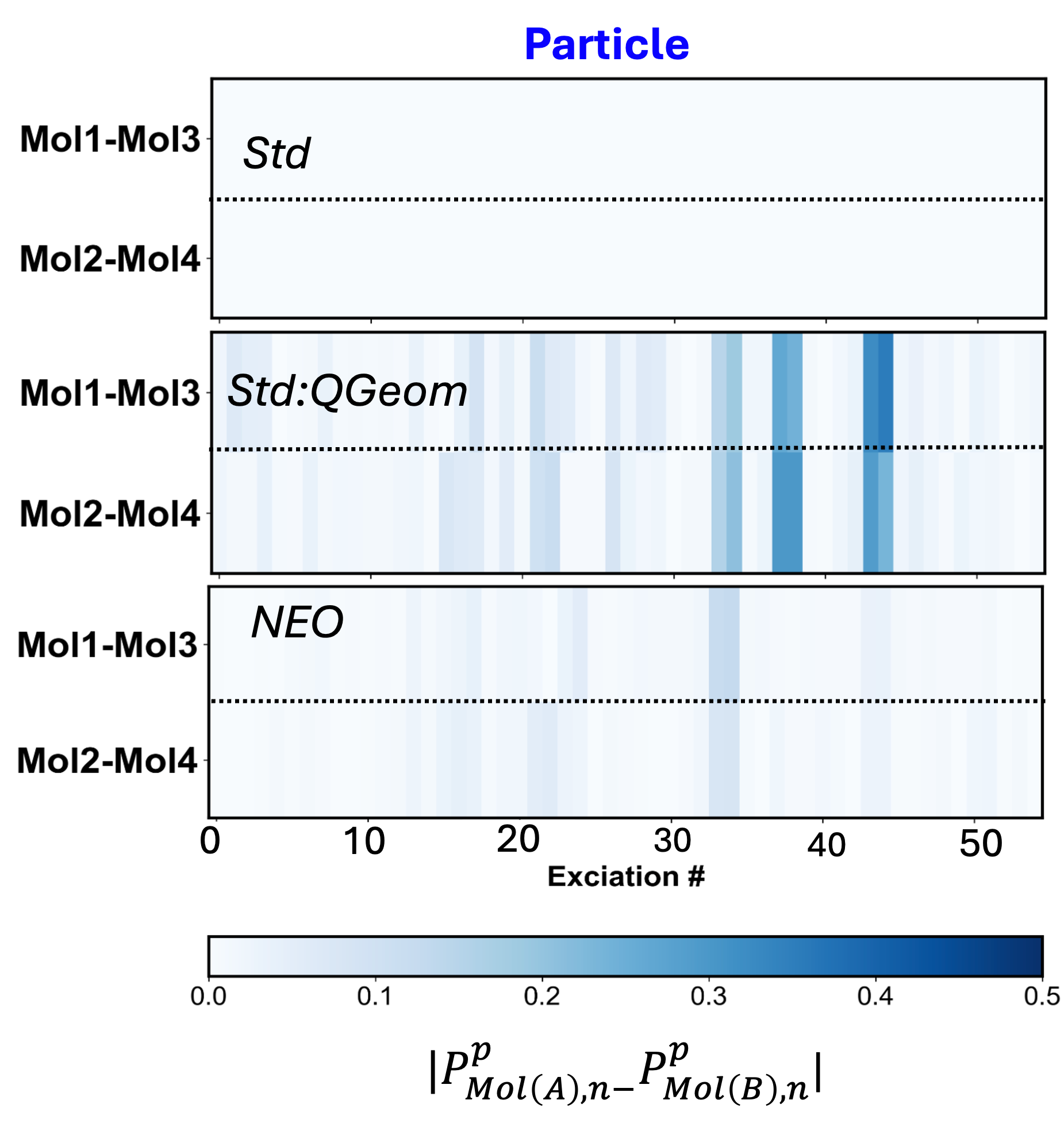}}
\subfigure[]{\label{fig:P_mol_diff_p}\includegraphics[width=70mm,height=80mm]{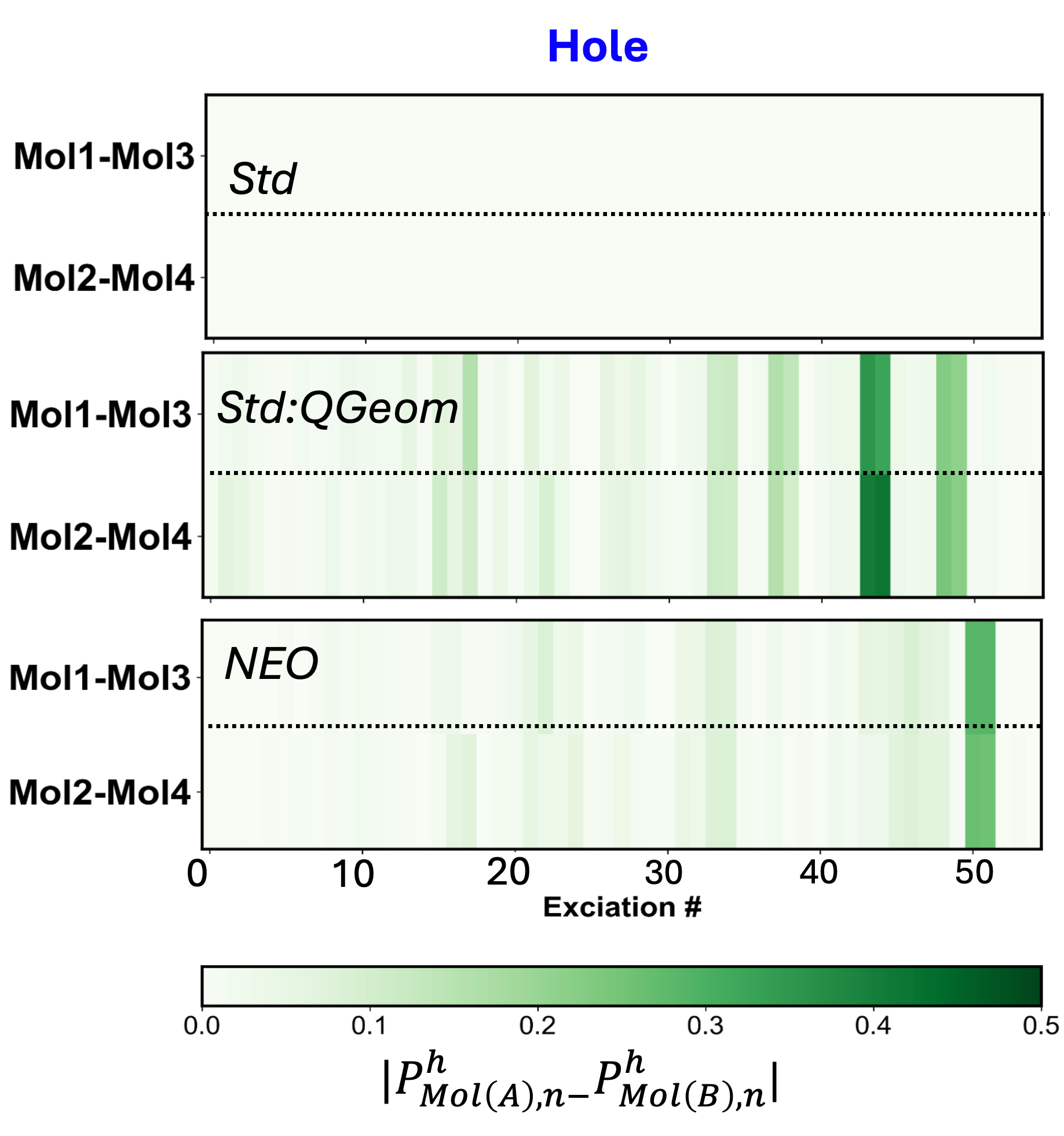}}
\caption{
The spatial anisotropy of excitons for hydrogen-bonded monomer units is quantified by
$|P^{p/h}_{Mol(A), n}-P^{p/h}_{Mol(B), n}|$ where
$A$ and $B$ are hydrogen-bonded pairs, for (a) particle and (b) hole.
The dark shades are indicative of the exciton anisotropy.
Both figures show the results for \textit{Std}, \textit{Std:QGeom}, and \textit{NEO} calculations.
Horizontal axes show the excitation number for the first 55 excitations, and the results for the higher excited states are provided in Figure S5 in the Supporting Information. 
}
\end{figure}
However, with quantized protons, much broader distributions are observed on all monomers, with the standard deviations two orders of magnitude larger. 
The excitons tend to be more anisotropic across different monomers. 
The proton quantum effects clearly impact the spatial extent of the exciton probability densities with significant anisotropy in the excited states, while such an effect is too subtle to show up in the optical absorption spectrum. Note that the \textit{Std:QGeom} calculation (Figure \ref{fig:sd_delocalization}) also shows the anisotropy, but close inspection shows notable differences, indicating that the proton quantum effects are not purely geometric effects in nature. 

Similarly to the work by \citeauthor{sasikumar2022exciton}\cite{sasikumar2022exciton}, hydrogen bonds have been proposed to impact exciton delocalization in other organic solids, such as organic light-emitting diodes\cite{sun2020charge} and supramolecular aggregates \cite{eric2023manifestation}.
In the unit cell of DHI, monomer pairs of Mol1-Mol3 and Mol2-Mol4 are hydrogen-bonded as shown in Figure \ref{fig:geom_1}.
To quantify the anisotropy of the exciton density in hydrogen-bonded pairs, the differences in the Mulliken exciton populations in these pairs are shown in Figures \ref{fig:P_mol_diff_el} and \ref{fig:P_mol_diff_p} for the particle and hole, respectively (i.e. $|P^{p/h}_{Mol(A), n}-P^{p/h}_{Mol(B), n}|$).
When the protons are not quantized, the exciton probability densities are essentially homogeneous. 
With the proton quantization, some anisotropy is observed for the hydrogen-bonded pairs, particularly for the 51st excited state.
Notably, this is not purely a geometry-derived effect of the proton quantization as can be deduced from comparing the \textit{NEO} and \textit{Std:QGeom} calculations.
\begin{figure}[htbp]
\captionsetup{justification= centerlast}
\centering
 \subfigure[\textit{Std}-1$^{\mathrm{st}}$]{\label{fig:std_ex_1}\includegraphics[width=73mm,height=50mm]{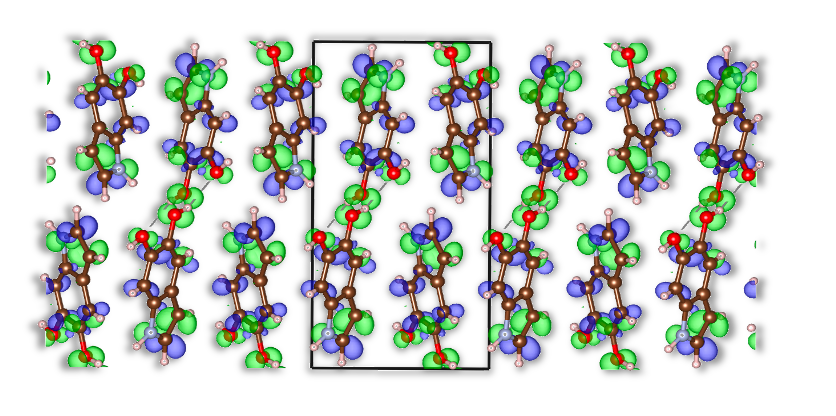}}
\subfigure[\textit{NEO}-1$^{\mathrm{st}}$]{\label{fig:neo_ex_1}\includegraphics[width=73mm,height=50mm]{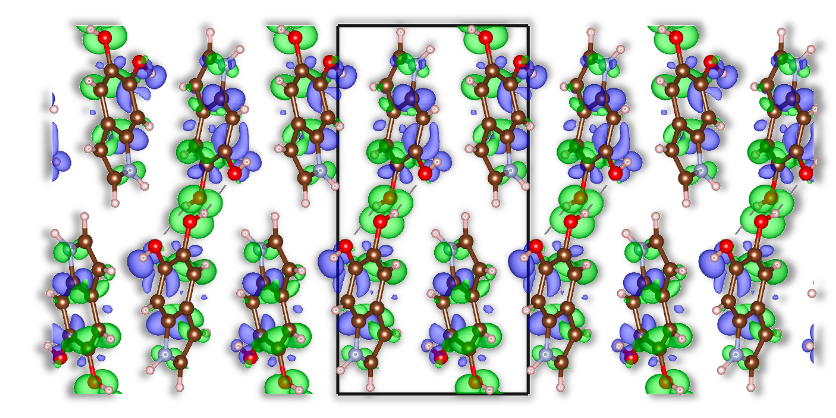}}
 \subfigure[\textit{Std}-51$^{\mathrm{st}}$]{\label{fig:std_ex_51}\includegraphics[width=73mm,height=50mm]{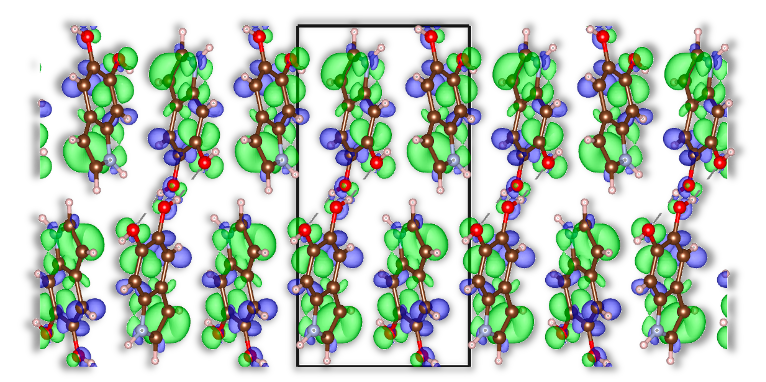}}
 \subfigure[\textit{NEO}-51$^{\mathrm{st}}$]{\label{fig:neo_ex_51}\includegraphics[width=73mm,height=50mm]{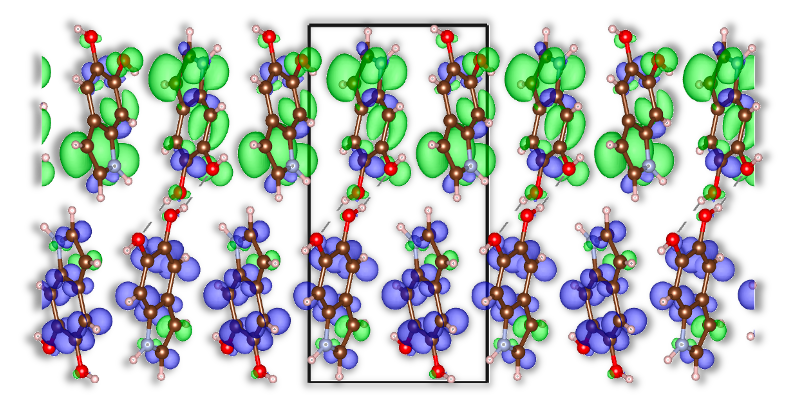}}
\caption{
Exciton particle and hole probability densities are shown for the 1$^{\mathrm{st}}$ and 51$^{\mathrm{st}}$ excited states. The corresponding excitation energies are $E_1^{\mathrm{Std}} = 4.49$ eV and $E_1^{\mathrm{NEO}} = 4.48$ eV, and $E_{51}^{\mathrm{Std}} = 6.14$ eV and $E_{51}^{\mathrm{NEO}} = 6.09$ eV. Particle and hole densities are shown in blue and green, respectively. Panels (a,c) correspond to \textit{Std} calculations, while panels (b,d) correspond to \textit{NEO} calculations. The black box indicate the unit cell.
}
\label{fig:excited_state_cube}
\end{figure}
Figure \ref{fig:std_ex_1} and \ref{fig:neo_ex_1} show the exciton probability density in the first excited state for the \textit{Std} calculation (E$_{1}^{Std}=$ 4.49 eV) and the calculation with the proton quantization (E$_{1}^{NEO}=$ 4.48 eV), respectively. 
Note that, in many theoretical works on excitons, the exciton probability density is plotted by fixing either the particle or hole coordinate at a particular point, making it look as if the exciton is localized around the particular point\cite{dai2024identification,dai2024theory,sharifzadeh2013low}.
The particle and hole of the exciton are homogeneously distributed over the four monomer units in the unit cell of the crystal while they tend to localize on different atoms within each monomer. 
As an exception to this general observation that holds also for other excited states, the exciton probability density for the 51$^{\mathrm{st}}$ excited state (E$_{51}^{Std}=$ 6.14 eV and E$_{51}^{NEO}=$6.09 eV) is shown in Figure~\ref{fig:std_ex_51} and Figure~\ref{fig:neo_ex_51}.
For this particular excited state, the proton quantum effects yield pronounced changes in its exciton distribution.
Without the proton quantization, both particle and hole densities of the exciton are delocalized across all four monomer units as can be seen in Figure~\ref{fig:std_ex_51}.

However, with the protons quantized, as in \textit{Std:QGeom} or \textit{NEO} calculations,  hydrogen atoms are no longer in their high symmetry positions. The structural symmetry breaking tends to result in significant variations in the Mulliken exciton population distribution across individual monomer units, for example, the hole density (green) of the exciton becomes localized over the two hydrogen-bonded monomer units as seen in Figure~\ref{fig:neo_ex_51} while the particle density (blue) remains isotropically delocalized. 
Similar localization patterns were observed for the \textit{Std:QGeom} exciton densities, and they are shown in Figure S5 in the Supporting Information. 
While the proton positions in both \textit{NEO} and \textit{Std:QGeom} calculations deviate from high symmetry sites, 
the delocalization of quantum protons is only present in the \textit{NEO} case and it effectively ``smooths out" the
effective potential electrons are subjected to. 
This leads to a more homogeneous electronic structure and thus a narrower distribution of Mulliken exciton populations in \textit{NEO} compared to \textit{Std:Qgeom}. 
We note here that the exciton features cannot be
compared between the calculations (i.e. \textit{NEO} and \textit{Std}) using the same excitation number on a one-to-one basis because the underlying ground-state electronic structures are not the same.


Despite their crucial role in accurately describing the structure of hydrogen-bonded condensed-phase systems such as water, the importance of nuclear quantum effects for electronic excitations remains largely unexplored.
Hydrogen bonds have been proposed to mediate exciton delocalization in the prototypical hydrogen-bonded organic solid of eumelanin\cite{sasikumar2022exciton}, and we studied how the quantization of protons impact electronic excitation using first-principles Green's function theory in this work. 
Employing the nuclear-electronic orbital (NEO) method within the Bethe-Salpeter equation (BSE) based on the $GW$ approximation,
we discussed the extent to which the electronic excitations 
are changed when the quantum nature of protons is taken into account.  
We also discussed the extent to
which the proton quantum effects arise from geometry-derived changes and
how they induce molecular-level anisotropy for excitons in the organic solid.
For the optical absorption spectrum, proton quantum effects were found to alter peak heights somewhat, although not significantly. 
As for the excitation energies, the proton quantization change them up to approximately 0.1 eV. 
Although not insignificant, the magnitude is smaller than typical errors one would expect from BSE@$GW$ method when calculating the excitation energies.

By adjusting the positions of classical protons according to the quantum proton position expectation values, the proton quantum effects in the optical absorption spectrum are qualitatively recovered.
However, on the scale of individual electronic excitations, the geometry-derived effects are not the only reason for the subtle molecular-level
exciton anisotropy that is induced by the quantization of protons. 
In summary, we demonstrated how the NEO method combined with the BSE@$GW$ theory can be used to study electronic excitations in organic solids by taking into account the proton quantum effects on their extensive hydrogen bonds. 
Organic solids tend to show strong exciton binding and some charge transfer character are often seen for the excitons. 
Our work here showed how the first-principles Green's function theory approach presented here will enable the community to examine nuclear quantum effects in electronic excitation in organic materials. Note that the possible correlation between electronic excitation and proton \textit{excitation} is neglected in this work. 
Overcoming this existing limitation in our method 
remains a daunting task, partly
due to the already large computational cost of BSE@$GW$ method, and it will be addressed in future work. 



\section{Computational/Theoretical Details}
Constrained geometry optimization was performed with a residual energy tolerance of 0.005 eV/atom by relaxing only hydrogen positions while the rest of the system is held fixed at the atom positions of the experimental XRD structure \cite{sasikumar2022exciton}.
Perdew-Burke-Ernzerhof (PBE) \cite{perdew1996generalized} generalized gradient approximation to the exchange-correlation (XC) functional was used for electrons,  combined with the Tkatchenko-Scheffler (TS) pairwise dispersion correction \cite{tkatchenko2009accurate}.
Given the flat bands (see Figure S1) in the Brillouin zone for this organic crystal, we used $\Gamma$-point only calculations in this work. 

In all calculations, the electronic wavefunctions were represented by the ``tier 2" numeric atomic orbital (NAO) basis set \cite{blum2009ab}. 
In the NEO-DFT framework, the proton wavefunctions were expanded using the PB4-D Gaussian-type basis sets\cite{yu2020development}, and the centers of these basis functions were iteratively optimized such that they coincide with the proton position expectation values. 
The proton-proton interactions are treated with the exact (Hartree-Fock) exchange, and the epc17-2 approximation was used for the electron-proton correlation (epc) \cite{brorsen2017multicomponent}. 
We emphasize that the resulting position expectation values of the quantum protons  from NEO-DFT calculations generally differ from the fixed nuclear positions of classical hydrogen atoms in standard DFT calculation.

$GW$ and BSE calculations were performed using the implementation in the FHI-aims code\cite{ren2012resolution,liu2020all,ren2021all,xu2022nuclear,bhattacharya2024bse,zhou2024all}.
We perform single-shot $G_0W_0$ to obtain quasi-particle energies, and the 
non-interacting Green function $G_0$ is constructed using Kohn-Sham eigenvectors and eigenvalues from DFT or NEO-DFT. The screened Coulomb interaction $W_0$ was computed within the random phase approximation (RPA) in the usual manner.
We included additional 4f spherical harmonic functions in the auxiliary basis to ensure convergence of the screened Coulomb operator within the RI-LVL scheme\cite{ihrig2015accurate} as discussed in Ref. \cite{zhou2024all}.
For the numerical evaluation of self-energy $\hat{\Sigma}$, we used the analytic continuation (AC) with the Pad\'e approximation \cite{van2015gw,liu2016cubic,wilhelm2018toward} as described in Ref. \cite{vidberg1977solving}. 
200 frequency points are utilized for integration on the imaginary axis. The BSE Hamiltonian was constructed using all the available states (147 valence and 44 conduction band states) within the Tamm-Dancoff approximation (TDA)\cite{hirata1999time}. 

To analyze the exciton in a particular excited state $n$, we utilize the exciton probability density $\rho_n(\mathbf{r_h},\mathbf{r_e})$, which represents the joint probability of finding the electron at position $\mathbf{r_e}$ and the hole at position $\mathbf{r_h}$ for a specific excited state $n$,

\begin{equation}
\rho_n(\mathbf{r_h},\mathbf{r_e})=\left |\sum_{vc\mathbf{k}}A_{vc\mathbf{k},n}\psi_{v\mathbf{k}}^*(\mathbf{r_h})\psi_{c\mathbf{k}}(\mathbf{r_e})\right |^2 
\label{eq:prob_density}
\end{equation}
where $A_{vc\mathbf{k},n}$ is the eigenvector obtained 
by solving BSE matrix 
(see Equation 11 in the  Supporting Information). $c$ and $v$ are conduction and valence state indices,  respectively. 
For deriving more 
intuitive physical insights, one can define a ``mean" probability density by integrating out the positional degrees of freedom associated with the other quasiparticle for the exciton (particle-hole pair), 
\begin{equation}   
\bar{\rho}_n^{h}(\mathbf{r_h})=\int \rho_n(\mathbf{r_h},\mathbf{r_e}) d\mathbf{r_e}
\label{eq:h_density}
\end{equation}
\begin{equation}   
\bar{\rho}_n^{p}(\mathbf{r_e})=\int \rho_n(\mathbf{r_h},\mathbf{r_e}) d\mathbf{r_h}.
\label{eq:e_density}
\end{equation}
\noindent
Here, the above ``mean probability density" can be conceptualized as an extension of the well-established detachment/attachment density analysis \cite{head1995analysis,dreuw2005single}—a method frequently used for analyzing molecular excited state wavefunctions—to extended periodic systems. 
To quantify the spatial distribution of the exciton and to analyze the contribution of each atom to a given excited state $n$, we introduce an atom-specific Mulliken population analysis. The Mulliken exciton population on a specific atom $N$ for the $n$-th excited state is defined for the hole and particle components of the exciton probability density separately.
The Mulliken population for the hole on atom $N$ in the excited state $n$ is given by
\begin{equation}
    P_{N, n}^{h} = \sum_{\mu \in N} \sum_{\nu} \sum_{\mathbf{R}} \left(D_{n}^{h}\right)_{\mu\mathbf{R}, \nu\mathbf{0}} S_{\mu\mathbf{R},\nu\mathbf{0}}
    \label{eq:mul_hole}
\end{equation}
Similarly, the Mulliken population for the particle on atom $N$ is
\begin{equation}
    P_{N, n}^{p} = \sum_{\mu \in N} \sum_{\nu} \sum_{\mathbf{R}} \left(D_{n}^{p}\right)_{\mu\mathbf{R}, \nu\mathbf{0}} S_{\mu\mathbf{R}, \nu\mathbf{0}}
    \label{eq:mul_particle}
\end{equation}
where the summation for $\mu$ runs over all atomic orbital basis functions centered on the particular atom $N$, while the summation for $\nu$ extends over all atomic orbital basis functions in the reference unit cell labeled by $\mathbf{0}$. $\mathbf{R}$ denotes the lattice vector of the periodic cell, and $S_{\mu\mathbf{R},\nu\mathbf{0}}$ represents the overlap matrix between atomic orbital basis functions with periodic boundary conditions. 
The terms $\left(D_{n}^{h}\right)_{\mu\mathbf{R}, \nu\mathbf{0}}$ and $\left(D_{n}^{p}\right)_{\mu\mathbf{R}, \nu\mathbf{0}}$ are the hole and particle density matrices, respectively, expressed in the basis of numeric atomic orbitals \cite{blum2009ab}. These density matrices are related to the
density matrices in the molecular orbital (MO) basis via a linear transformation 
\begin{equation}
\left(D_{n}^{h}\right)_{\mu\mathbf{R}, \nu\mathbf{0}} = \sum_{v_1, v_2} \sum_{\mathbf{k}} e^{i\mathbf{k} \cdot \mathbf{R}} \left(D_{n}^{h}\right)_{v_1 v_2, \mathbf{k}} c_{\mu, v_1\mathbf{k}} c_{\nu, v_2\mathbf{k}}^{*}
    \label{eq:density_matrix_hole_ao}
\end{equation}

\begin{equation}
    \left(D_{n}^{p}\right)_{\mu\mathbf{R}, \nu\mathbf{0}} = \sum_{c_1, c_2} \sum_{\mathbf{k}} e^{i\mathbf{k} \cdot \mathbf{R}} \left(D_{n}^{p}\right)_{c_1 c_2, \mathbf{k}} c_{\mu, c_1\mathbf{k}} c_{\nu, c_2\mathbf{k}}^{*}
    \label{eq:density_matrix_particle_ao}
\end{equation}
where $c_{\mu, v\mathbf{k}}$ and $c_{\mu, c\mathbf{k}}$ are the coefficients for the atomic orbital $\mu$ in the Bloch wavefunctions of the valence band ($v$) and conduction band ($c$) states, respectively, at a given k-point $\mathbf{k}$ in the Brillouin zone.
The density matrices in the MO basis, $\left(D_{n}^{h}\right)_{v_1 v_2, \mathbf{k}}$ and $\left(D_{n}^{p}\right)_{c_1 c_2, \mathbf{k}}$, are constructed from the exciton amplitudes, $A_{vc\mathbf{k}, n}$, which are the solutions of the Bethe-Salpeter equation for the $n$-th exciton. The {particle/hole density matrix} in the MO basis is given by
\begin{equation}
    \left(D_{n}^{p}\right)_{c_1 c_2, \mathbf{k}} = \sum_{v} A_{v c_1 \mathbf{k}, n}^{*} A_{v c_2 \mathbf{k}, n}
    \label{eq:density_matrix_particle_mo}
\end{equation}
\begin{equation}
    \left(D_{n}^{h}\right)_{v_1 v_2, \mathbf{k}} = \sum_{c} A_{v_1 c \mathbf{k}, n}^{*} A_{v_2 c \mathbf{k}, n}
    \label{eq:density_matrix_hole_mo}
\end{equation}
Here, the superscripts $p$ and $h$ distinguish between the particle (electron) and the hole of the $n$-th exciton, as seen in equations \eqref{eq:mul_hole} through \eqref{eq:density_matrix_hole_mo}. The indices $c, c_1, c_2$ denote conduction band states while $v, v_1, v_2$ refer to valence band states. The summation over a single k-point $\mathbf{k}$ is a consequence of the momentum conservation in the optically induced formation of the particle-hole pair as an exciton in an extended system.

To examine the extent to which the exciton probability density is delocalized for a particular excited state, $n$, we further define the Mulliken exciton population for each individual monomer fragment ($Mol$) in the unit cell 
\begin{align}
P_{{Mol}, n}^{p/h} & =\sum_{N\in {Mol}}  P_{N, n}^{p/h}. \hspace{5mm}
\label{eq:Mol_decom}
\end{align}
where $\sum_{N\in {Mol}}$ denotes the summation over those atoms $N$ which belong to the monomer fragment $Mol$ (see Eq. \ref{eq:mul_hole} and \ref{eq:mul_particle}).
For convenience, one gets the standard deviation of these Mulliken exciton populations for the manifold of excited states:
\begin{align}
\sigma(P^{p/h}_{Mol}) & =\sqrt{\frac{1}{n_{max}}\sum_n^{n_{max}}(P^{p/h}_{Mol, n}-\mu_{P_{Mol}}^{p/h})^2}
\label{eq:spread}
\end{align}
where $n_{max}$ is the maximum number of the excited states included in the manifold and $\mu_P$ is the average value of $P^{p/h}_{Mol, n}$.

\section*{Supporting Information}



Supporting Information includes computational details, theoretical framework of NEO-DFT, all-electron periodic $G_0W_0$, and Bethe-Salpeter Equation formalisms,
as well as additional data on density of states, oscillator strengths, and spatial anisotropy of excitons for higher excited states.

\bibliography{scibib}

\end{document}